\documentclass[10pt,conference]{IEEEtran} 
\IEEEoverridecommandlockouts

\usepackage[numbers]{natbib}
\usepackage{booktabs}
\usepackage{multirow,multicol}
\usepackage{amsmath,amssymb,amsfonts}
\usepackage{algorithmic}
\usepackage{graphicx}
\usepackage{textcomp}
\usepackage{xcolor}
\usepackage[colorlinks=false,hidelinks=true]{hyperref}

\usepackage[most]{tcolorbox}
\usepackage{soul}

\def\BibTeX{{\rm B\kern-.05em{\sc i\kern-.025em b}\kern-.08em
    T\kern-.1667em\lower.7ex\hbox{E}\kern-.125emX}}
\begin{document}

\title{Syntax and Domain Aware Model for  \\ Unsupervised Program Translation}

\author{\IEEEauthorblockN{Fang Liu}
\IEEEauthorblockA{State Key Laboratory of Software\\Development Environment \\
Beihang University\\
Beijing, China \\
fangliu@buaa.edu.cn}
\and
\IEEEauthorblockN{Jia Li}
\IEEEauthorblockA{Key Lab of High Confidence\\Software Technology, MoE \\
Peking University\\
Beijing, China \\
lijiaa@pku.edu.cn}
\and
\IEEEauthorblockN{Li Zhang\IEEEauthorrefmark{1}}
\IEEEcompsocitemizethanks{\IEEEcompsocthanksitem\IEEEauthorrefmark{1}Corresponding author}
\IEEEauthorblockA{State Key Laboratory of Software\\Development Environment \\
Beihang University\\
Beijing, China \\
lily@buaa.edu.cn}

}

\maketitle

\begin{abstract}
There is growing interest in software migration as the development of software and society. Manually migrating projects between languages is error-prone and expensive. In recent years, researchers have begun to explore automatic program translation using supervised deep learning techniques by learning from large-scale parallel code corpus. However, parallel resources are scarce in the programming language domain, and it is costly to collect bilingual data manually. To address this issue, several unsupervised programming translation systems are proposed. However, these systems still rely on huge monolingual source code to train, which is very expensive. Besides, these models cannot perform well for translating the languages that are not seen during the pre-training procedure. In this paper, we propose SDA-Trans, a syntax and domain-aware model for program translation, which leverages the syntax structure and domain knowledge to enhance the cross-lingual transfer ability. SDA-Trans adopts unsupervised training on a smaller-scale corpus, including Python and Java monolingual programs. The experimental results on function translation tasks between Python, Java, and C++ show that SDA-Trans outperforms many large-scale pre-trained models, especially for unseen language translation. 
\end{abstract}


\section{Introduction}
Programming language is the main tool for creating desktop applications, websites, and mobile applications. 
To develop different applications, many different programming languages have been invented over the years. The diverse programming languages can facilitate developers to build various applications of multiple environments and platforms.
To meet new business needs and environments, one software initially developed in a language needs to migrate to other languages \cite{Wu2010AURA,nguyen2015divide}. Code migration is a crucial activity during software development and maintenance \cite{chen2018tree}. There is a tremendous need for translating programs between different programming languages.

Early studies mainly relied on handcrafted translation rules to translate between two languages. The translation is poor in readability and correctness, and needs extra manual corrections. Therefore, the translation process is error-prone and time-consuming \cite{Zhong2010miningAPI}.
For example, it takes about 5 years and \$750 million for Australia Commonwealth Bank to convert the platform from COBOL to Java \cite{TranCoder2020}. 
To improve the quality and efficiency, researchers began to explore the automatic approaches to build a program translator \cite{nguyen2013lexical,nguyen2015divide,chen2018tree,TranCoder2020}. Based on the naturalness theory of programming languages, i.e., source code is repetitive and predictable, statistical machine learning methods have been used for modeling programs and code migration. Some previous work borrowed the idea from the natural language translation field and applied statistical machine translation (SMT) approaches \cite{Koehn2010SMT} for code migration \cite{nguyen2013lexical,nguyen2015divide}. Among these studies, researchers build statistical models and optimize the parameters by analyzing the statistical relationships between bilingual data. Then they employ these statistical models to generate translations. 

With the development of deep learning technologies, in recent years, researchers have started to tackle program translation task using supervised sequence-to-sequence neural machine translation (NMT) methods \cite{Bahdanau2014NMT}, which need a large amount of parallel code resources to train. However, parallel resources are much more scarce in the programming language domain than in natural language.
It is costly to collect bilingual program data manually. Therefore, applying the NMT technology to code translation still faces many challenges. To ease this issue, unsupervised machine translation techniques are proposed for programming translation \cite{TranCoder2020,TransCoderST2022}. TransCoder \cite{TranCoder2020} first applied unsupervised machine translation approaches for program translation task, leveraging huge monolingual programs from GitHub to train their model, covering C++, Java, and Python languages. Then they adopted the model to translate between C++, Java, and Python. However, the model still needs to be pre-trained using a large amount of monolingual code corpus, which is very expensive. Besides, it cannot fit well for the languages that are not seen during the pre-training procedure according to our experimental results.

\begin{figure*}[t]
    \centering
    \setlength{\abovecaptionskip}{0.1cm} 
    \includegraphics[width=0.7\linewidth]{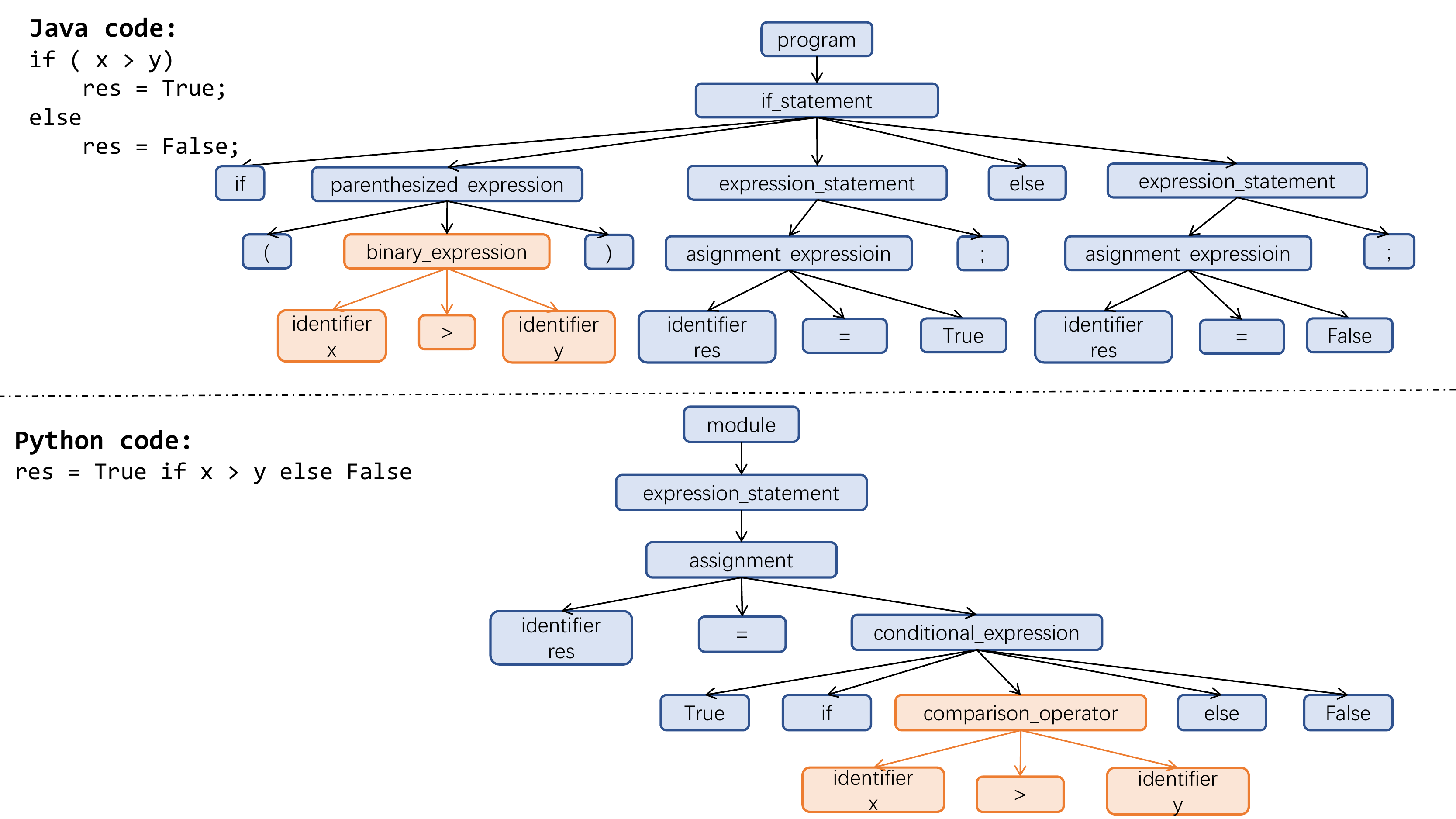}
    \caption{If statements in Java and Python and their corresponding ASTs.}
    \label{fig:motivating_example}
    \vspace{-0.3cm}
\end{figure*}

Cross-lingual transfer can reduce the demand for labeled data to perform corresponding tasks in a new language and thus can support applications in low-resource languages \cite{Ahmad2020SynBERT}. We argue that facilitating cross-lingual transfer is an effective way to improve data utilization and save computing resources, which can benefit program translation systems.
Transferring across programming languages is challenging due to the differences in morphology, syntax, and semantics. 
For instance, the \textit{IF} statements in Python and Java in Figure \ref{fig:motivating_example} have the same meaning but with different token orders.
Unlike natural languages, programs have explicit syntax structure and rigorous grammar. The correlation between syntax structure is much closer, which can help to reduce the morphological differences across languages and extract language-invariant features. As seen from Figure \ref{fig:motivating_example}, the sub-trees corresponding to $x \textgreater y$ (highlighted in the figure) in the two ASTs have the same structure. Besides, the distance between node pairs can indicate the structural correlation between tokens, and further reflects specific patterns of code snippets. 

However, the syntax structure is not fully leveraged in existing program translation models \cite{TranCoder2020,TransCoderST2022}. 
Besides, these models are pre-trained with several self-supervised tasks on specific code corpus. Thus, they are poor at distinguishing the characteristic of different languages and domains when transferring knowledge.
To this end, we proposed \textbf{SDA-Trans}, a \textbf{S}yntax and \textbf{D}omain-\textbf{A}ware model for program \textbf{Trans}lation. Specifically, we augment Transformer model with language syntax by incorporating the structured representations of the program into the self-attention mechanism. Besides, we design a new domain-distinguish pre-training task to extract domain-invariant features\footnote{The definition of the domain-invariant features can be found in Section \ref{Preliminaries}.}. This task asks the model to distinguish which domain/language the model is translated from, and learn the task via adversarial training. This task will encourage the model to distill domain-invariant syntactic and semantic features.
Inspired by unsupervised machine translation approaches \cite{Lample2018Phrase-BasedNMT,TranCoder2020}, we train SDA-Trans with the following steps: (1) Initializing parameters of the vanilla Transformer with Cross Programming Masked Language Model and Denoising Auto-Encoding tasks; (2) Augmenting the model with syntax information, and add domain-distinguish pre-training task to make our model domain-aware by adversarial training; (3) Tuning SDA-Trans with back-translation. 

Since SDA-Trans is designed to extract the syntax structure and domain-invariant features during pre-training, we hope it can learn enough knowledge from limited data, which can, in turn, save computing resources. We train our model on a smaller scale of Python and Java monolingual code corpus (which contains 165k Java methods and 251k Python methods) compared with existing large-scale pre-trained models. We evaluate SDA-Trans on program translation between Java, Python, and C++, where C++ language is never seen during pre-training. Experimental results show that SDA-Trans achieves a remarkable gain compared with the large-scale pre-trained unsupervised program translation methods.
The ablation study further shows that both syntax and domain augmentation contribute to the improvement. 
We summarize the contributions of this paper as follows:
\begin{itemize}
    \item We build a novel approach SDA-Trans for unsupervised program translation, which leverages the syntax structure and domain knowledge to enhance the model's cross-lingual transfer ability.
    \item We present a new unsupervised training paradigm using a three-stage strategy: (1) initializing the model with CMLM and DAE objectives; (2) continuously training with syntax and domain augmentation; and (3) tuning with back-translation.
    \item Despite pre-trained with fewer data, SDA-Trans achieves impressive performance on program translation, which is comparable with the large-scale pre-trained models, especially on unseen language translation.
\end{itemize}

\section{Background \& Research Question Formulation}

This section first describes the background knowledge that is related to this work, and then gives an overview of the research questions.

\subsection{Supervised Machine Translation for Program Translation}
From the probabilistic perspective, the translation's purpose is to find a target code $y$ with the maximum conditional probability given the source code $x$.
Statistical Machine Translation (SMT) \cite{Koehn2010SMT} approaches first train models to learn the translation rules from bilingual code corpora. Then they perform translation based on the trained model.

Later, seq2seq models based on deep neural networks dominated the translation task in the natural language processing (NLP) field.
In neural machine translation (NMT) \cite{Bahdanau2014NMT}, a parameterized encoder-decoder network is typically used. The parameters are optimized via maximizing the conditional probability of the parallel statement pairs in the data corpus. Once the translation model is well-fitted for learning the conditional distribution given a source sentence, the model can generate the corresponding translation with a maximized conditional probability. The most widely used models for NMT pertain to a family of encoder-decoder architectures with attention mechanism \cite{Bahdanau2014NMT}, where the encoder is responsible for encoding a source sentence, and a decoder generates the translation based on the encoder output and the attention mechanism. RNN and its variants are generally used to instantiate the encoder and decoder.

Since the self-attention mechanism \cite{vaswani2017attention} is proposed, Transformer-based models have achieved extraordinary performance in machine translation.

\subsection{Unsupervised Machine Translation For Program Translation}

NMT systems' performance strongly depends on the quality and quantity of the parallel data corpus. In practice, parallel resources are rare and costly to obtain for most languages. Researchers began investigating unsupervised and weakly supervised machine translation techniques to address this issue. Some approaches tried to use monolingual code corpus to improve the translation performance. \citet{he2016dual} built a translation model to learn from unlabeled data through dual learning. Later, \citet{Artetxe2018UnsupervisedNMT} proposed a novel unsupervised NMT model, which is trained solely on monolingual corpora. They first employ a shared encoder to train the system for input reconstruction task with monolingual data. Then to boost the translation performance, they introduce a back-translation technology into the training procedure. In the same year, \citet{Gu2018UniversalNMT} designed a universal machine translation approach, which adopted the transfer learning method to share the representation of the lexical level and sentence level of various languages, thus can further enable resource sharing between the languages of high resource and extremely low resource.

\subsection{Research Questions}

We hypothesize that explicitly providing language syntax constraints and domain knowledge can help to bridge the morphological and topological gaps across programming languages, and endow the model with the ability to learn domain-invariant representations among different languages, thus can bring benefits for cross-lingual transfer and program translation. This leads to several RQs.

\begin{tcolorbox}[enhanced,colback=gray!5!white,colframe=gray!75!black,drop fuzzy shadow southwest,fontupper=\bfseries]
  RQ1. How effective is SDA-Trans for program translation? 
\end{tcolorbox}
SDA-Trans uses a small-scale training corpus in contrast to existing large-scale pre-trained techniques \cite{TranCoder2020,CodeT52021}. 
We investigate whether leveraging the syntax and domain information can help our model learn sufficient knowledge from a limited amount of data and compare our SDA-Trans with several advanced program translation approaches.

\begin{tcolorbox}[enhanced,colback=gray!5!white,colframe=gray!75!black,drop fuzzy shadow southwest,fontupper=\bfseries]
  RQ2. How do different components in SDA-Trans contribute to performance improvement?
\end{tcolorbox}
We performed an ablation study to figure out the contribution and effectiveness of each proposed component in SDA-Trans.  

\begin{tcolorbox}[enhanced,colback=gray!5!white,colframe=gray!75!black,drop fuzzy shadow southwest,fontupper=\bfseries]
  RQ3. How well does our approach perform in unseen language translation scenarios? 
\end{tcolorbox}
To examine whether augmenting Transformer with syntax and domain knowledge improves the cross-lingual transfer ability, we further perform experiments on C++ related translation tasks. Since C++ language is unseen during the pre-training procedure of SDA-Trans, the results can reflect our model's performance in unseen language translation scenarios.

\begin{tcolorbox}[enhanced,colback=gray!5!white,colframe=gray!75!black,drop fuzzy shadow southwest,fontupper=\bfseries]
  RQ4. How useful are the programs translated by SDA-Trans in practice?  
\end{tcolorbox}
To analyze the semantic correctness and usefulness of the programs translated by SDA-Trans, we use computational accuracy {\cite{TranCoder2020}} to evaluate whether the programs produced by SDA-Trans can generate the same outputs as the ground truth given the same input data.

\section{SDA-Trans}

SDA-Trans is a Transformer based model, which is syntax and domain aware by adding the relevant components. For code inputs, the encoder inputs the tokenized source code sequence. Meanwhile, we employ graph attention network (GAT) \cite{Velickovic2018GAT} to embed the syntax information of the input token sequence into a structured representation and then use it to enhance self-attention. The decoder is responsible for generating the target code sequence. In order to guide our model to learn the syntax structure and domain-agnostic knowledge, we adopt auxiliary objectives that supervise the model to learn the syntax structure and domain-invariant features.

\subsection{Preliminaries}\label{Preliminaries}

In this work, we focus on function-level program translation. Each code snippet can be tokenized into a sequence of (sub-) tokens\footnote{Following TransCoder, we use fastBPE to split tokens into sub-tokens.} $S = {s_1,s_2,...,s_{|S|}}$. A code $S$ has a corresponding syntax tree $T = (N, N_{leaf}, r, p(\cdot),M)$, which reflects the syntax structure of the code snippet. $N$ is the set of nodes in the tree, and $r \in N$ is the root. $N_{leaf} = \{l_1,l_2,...,l_{|N_{leaf}|}\} \subset N$ is the subset of leaf nodes, and each leaf node corresponds to each token of the code snippets. $p:N-R \to N$ indicates the relationship between parent and children, where $p(n)$ denotes the parent of node $n$. $M \in \{0,1\}^{|S| \times |N_{leaf}|}$ is a linking matrix, $M_{ij} = 1$ if (sub-) token $s_i$ is part of the leaf node $l_j$. 

Previous work has shown that the structural information of the AST can be well-captured by modeling the node distances and the node paths \cite{codetransformer2021,peng2021integrating}. Here we introduce two kinds of distances based on AST used in our model. 

\begin{figure}
    \centering
    \includegraphics[width=\linewidth]{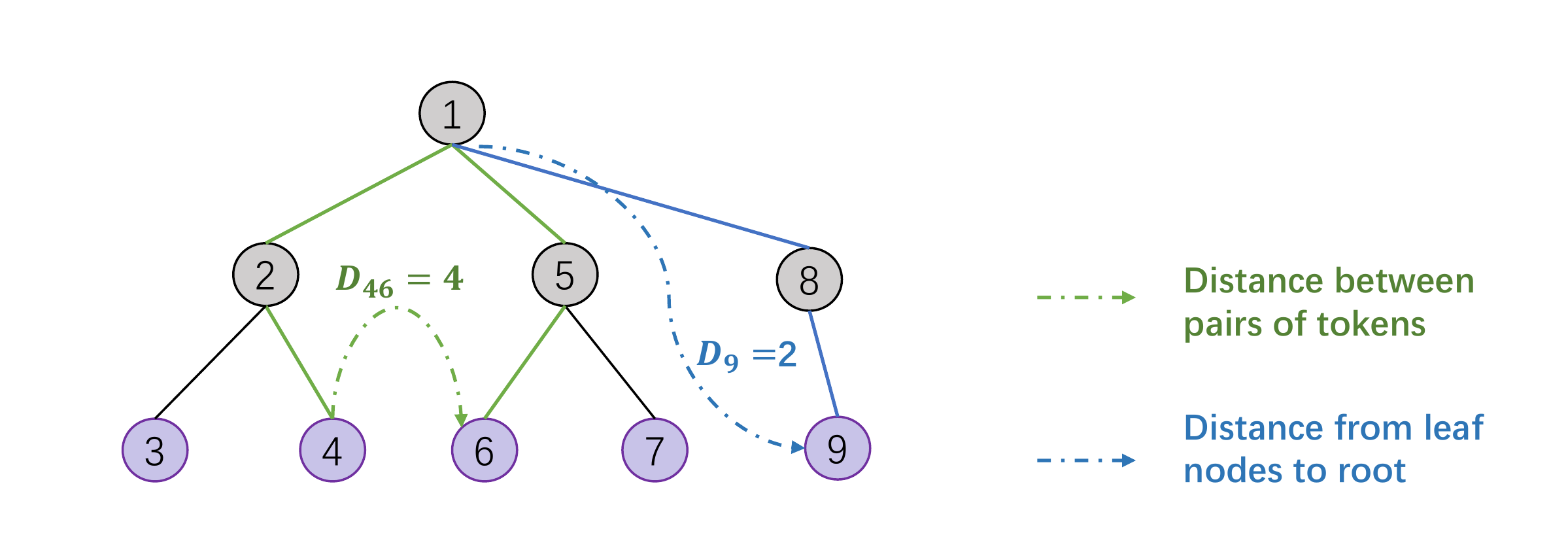}
    \caption{Structure distances used in SDA-Trans.}
    \label{fig:distance}
    \vspace{-0.3cm}
\end{figure}

\begin{itemize}
    \item Distance between pairs of tokens (leaf nodes) $D_{ij}$: this distance is the number of steps required to reach node $j$ from node $i$. We view the abstract syntax tree as an undirected graph. As shown in \ref{fig:distance}, the distance between node $4$ and node $6$ is 4.
    \item Distance from leaf nodes to root $D_{j}$: this distance depicts the tree depth of each leaf node (source code token) $j$, that is, the path length from node $j$ to root node. As shown in Figure \ref{fig:distance}, the distance from node $9$ to the root node is 2.
\end{itemize}

For program translation, transferring knowledge across programming languages is challenging due to domain discrepancies across languages. Thus, except for the syntax structural information, \textbf{domain-invariant} features are also crucial for enhancing domain generalization and cross-lingual transfer ability, which is an essential factor in program translation.
We consider the domain-invariant features as the features that are invariant under different languages, that is, the language features with semantic equivalence. Capturing the shared feature representation of different languages can lead to a good generalization performance on new target domains. 
The domain-invariant features can be reflected in many aspects, including syntax structure, data dependency relationship, and other semantic features. For example:\\
\noindent (1) A functionality or a specific syntax structure can be implemented differently for different languages. As shown in Figure {\ref{fig:motivating_example}}, the two if-else statements in Python and Java have the same meaning but with different token orders (one-line if-else is widely used in Python). The correlation between their syntax structure is much closer.\\
\noindent (2) The code snippets in various languages that implement the same functionality may look differently, however, the data dependency information among the variables will remain the same. For instance, the data flow relation between constants and variables, the call flow relation among statements and functions, \textit{etc}.\\
\noindent (3) Different symbols in different languages can have the same meaning. For instance, Python uses indentation to indicate a block of code, while Java and C++ use `\{' and `\}'. For the list append operation, Python uses ``append'' function, while Java uses ``add'' function.

To distill the complicated domain-invariant features as well as enhance the cross-lingual transfer ability, we build a model to learn the syntax structure explicitly, and also propose a domain-distinguishing task. Detailed information is presented in the following sections.

\subsection{Backbone Encoder}
SDA-Trans employs a vanilla Transformer encoder \cite{vaswani2017attention} as the base encoder. The encoder has several components, including an embedding layer and stacked multi-head self-attention blocks. 

\noindent\textbf{Embedding Layer}~ converts the input code sequence to embedding vectors, consisting of a token embedding matrix $W_e$ and a position embedding matrix $W_p$. Given the input code token sequence $w_1,w_2,...,w_n$ and their corresponding position sequence $p_1,p_2,...,p_n$, the embedding layer computes the output $H = \{x_1,x_2,...,x_n\} \in R^{n \times d_{model}}$, where $x_i = w_iW_e +p_iW_p$, $d_{model}$ is the encoder output dimension.

\noindent\textbf{Multi-head Self-Attention Block} is composed of $h$ attention heads, aiming at learning information from different representation sub-spaces. The output of each head is computed as follows \cite{vaswani2017attention}:

\begin{equation}
    \begin{split}
        Q = H^{l-1}W^Q_l \\
        K = H^{l-1}W^K_l \\
        V = H^{l-1}W^V_l
    \end{split}
\end{equation}
where $H_{l-1}$ is the output from the previous layer, $W^Q \in R^{d_{model} \times d_k}, W^K \in R^{d_{model} \times d_k}, W^V \in R^{d_{model} \times d_v}$ are the projection parameter matrices, which are used to linearly projecting $H_{l-1}$ into queries, keys, and values. 
Then scaled dot-product attention is applied to compute the output for each head:
\begin{equation}
    \begin{split}
        Attn(Q,K,V,M) = \text{softmax}(\frac{QK^T+M}{\sqrt{d_k}})V
    \end{split}
\end{equation}
where $d_k$ denotes the head dimension. $M \in R^{n \times n}$ is a mask matrix, which indicates whether two input tokens can see each other. Finally, the representations coming from all the attention heads are concatenated together and fed into a feed forward network to compute the final output: $H^l \in R^{n \times d_{model}}$.

\subsection{Syntax-augmented Encoder}\label{syntax_augmented_encoder}
To enhance the model's knowledge transfer ability and focus on learning the language invariant features, we augment the Transformer encoder with explicit syntax structure. Specifically, we use a graph attention network (GAT) \cite{Velickovic2018GAT} to encode the structural information of the code snippet into vectors. The graph is built based on AST, where the nodes represent tokens (leaf nodes in the tree), and the edges represent token dependencies. Then we use the GAT outputs to bias the self-attention module. We detail each process as follows.

\begin{figure}
    \centering
    \includegraphics[width=\linewidth]{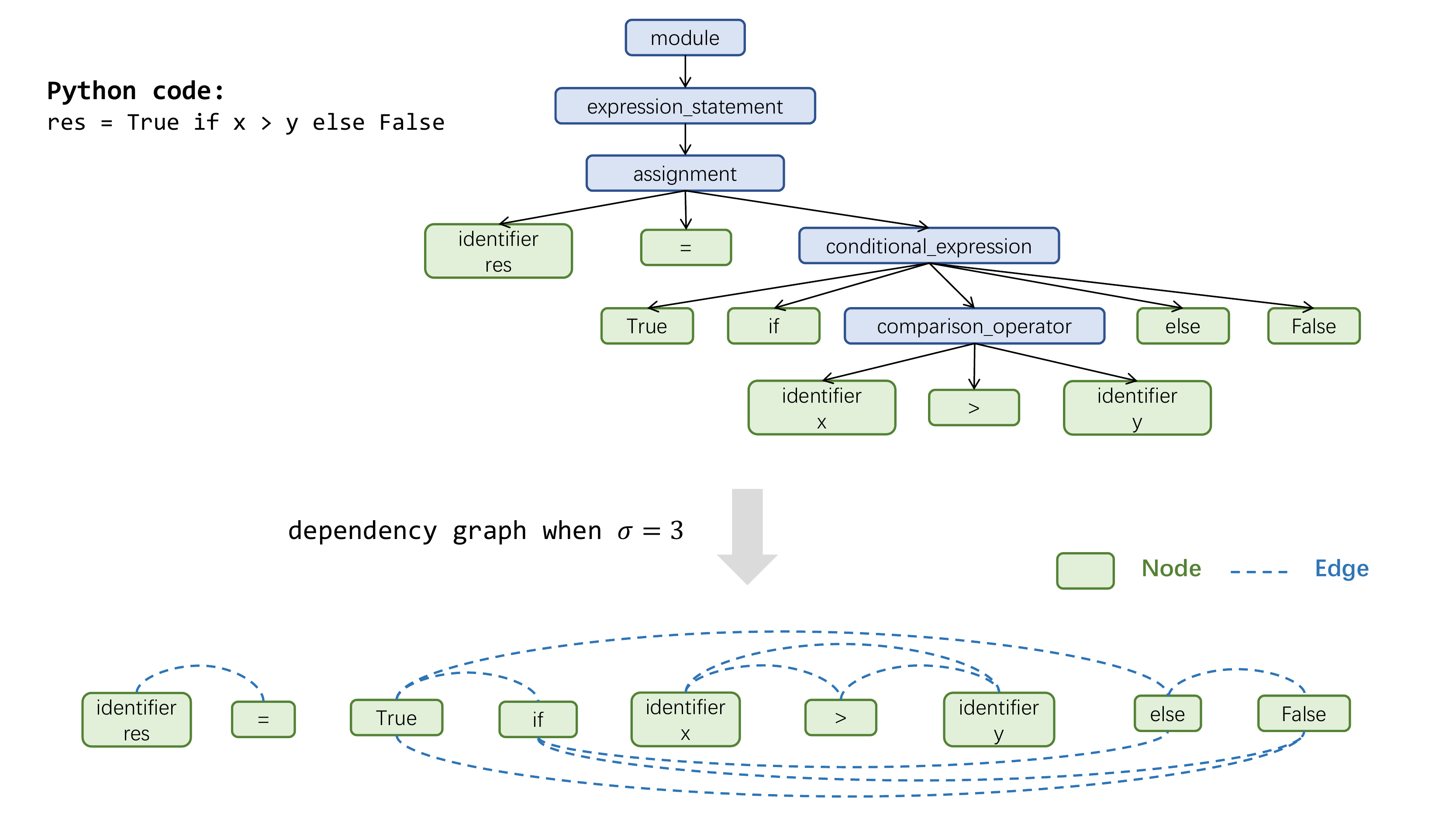}
    \caption{Dependency graph building process.}
    \label{fig:graph_build}
    \vspace{-0.3cm}
\end{figure}

\noindent\textbf{Dependency Graph Building} ~
The dependency graph is built based on AST, as shown in Figure \ref{fig:graph_build}. The original AST only includes the parent-children edges. We extract all the leaf nodes as the node set of the graph. Then we add edges for the node pairs whose distance is lower than a threshold $\sigma$, i.e., $D_{ij} \textless \sigma $. As we tokenize the input code sequence into subword units, the distance among the subword units of one token is set to 1, and the distance from the subword units to other nodes remains unchanged, which is equivalent to the shortest path length between their corresponding whole tokens. 

\noindent\textbf{GAT} ~ We use a GAT module to encode the tree structure of the code snippet. We first encode the input code into an embedding vector $\mathcal{G}^0$ using the same embedding matrix in the backbone encoder. Then we build a $L_{GAT}$-layer GAT to compute the structured output of each token from $\mathcal{G}^0$ by a dependency-aware self-attention mechanism. Specifically, each GAT layer is responsible for learning the token representation by attending to their adjacent tokens:

\begin{equation}
    O = Attention(Q,K,V,M)
\end{equation}
where Q=K, the mask $M$ is computed as:

\begin{equation}
M_{ij} = \left\{  
\begin{aligned}
    &0, \ \ \ \ \ \ D_{ij} < \sigma \\
    &-\infty, \ \text{otherwise}
\end{aligned}
\right.
\end{equation}
where $D_{ij}$ is the distance between token $i$ and token $j$ in the AST, as we have mentioned in section \ref{Preliminaries}.

Typically in the original GAT, $\sigma$ is set to 1, which only allows adjacent words to attend to each other. In our study, the graph is built based on the abstract syntax tree. The distance between node pairs is the number of steps required to reach from one node to another, which implies the structural correlation between code tokens and further reflects the specific patterns of the code snippets. Based on our preliminary experiments, $\sigma=10$ is the best setting. \footnote{The AST of code is much deeper than the parsing tree of the natural language, and only leaf nodes (corresponding to the source code tokens) are used as the final input.}
Finally, the output vectors of all the GAT are concatenated together and then used to produce the final output: $\mathcal{G}^l \in R^{n \times md_g}$, where $m$ is the number of GAT heads, $d_g$ is the head dimension.

\noindent\textbf{Structure-Aware Self-Attention}
To incorporate the structural syntax information into the model, we adopt a structure-aware self-attention, which considers the structural relations when computing attention scores between code tokens.
We use the representations produced by GAT to bias the self-attention:
\begin{equation}
    O = Attn(Q+\mathcal{G}G_Q^l,K+\mathcal{G}G_K^l,V,M)
\end{equation}
where $G_Q^l,G_K^l \in R^{d_{md_g} \times d_k}$ are the projection parameter matrices. $\mathcal{G}G_Q^l$ and $\mathcal{G}G_K^l$ are responsible for providing syntactic clues to guide self-attention, where each token can pay more attention to tokens with specific syntax dependencies.

Our backbone $L$-layer Transformer encoder employs $h$ attention heads. We upgrade part of these heads to \textit{GAT-heads}, which can fuse the syntax representations. Besides, we also upgrade several encoder layers to \textit{GAT-layers}, which can merge the syntax representations.

\subsection{Decoder}

To generate the target code snippets, we employ the vanilla Transformer decoder to sequentially predict the sub-token $y_t$ based on the previously generated sub-tokens $y_{\textless t}$ and the encoder output:
\begin{equation}
    p(y_t) = \text{softmax}(\text{FFN}(Wh_i))
\end{equation}
where $h_i$ is the decoder's hidden states.

\section{Training}

\begin{figure}[h]
    \centering
    \includegraphics[width=\linewidth]{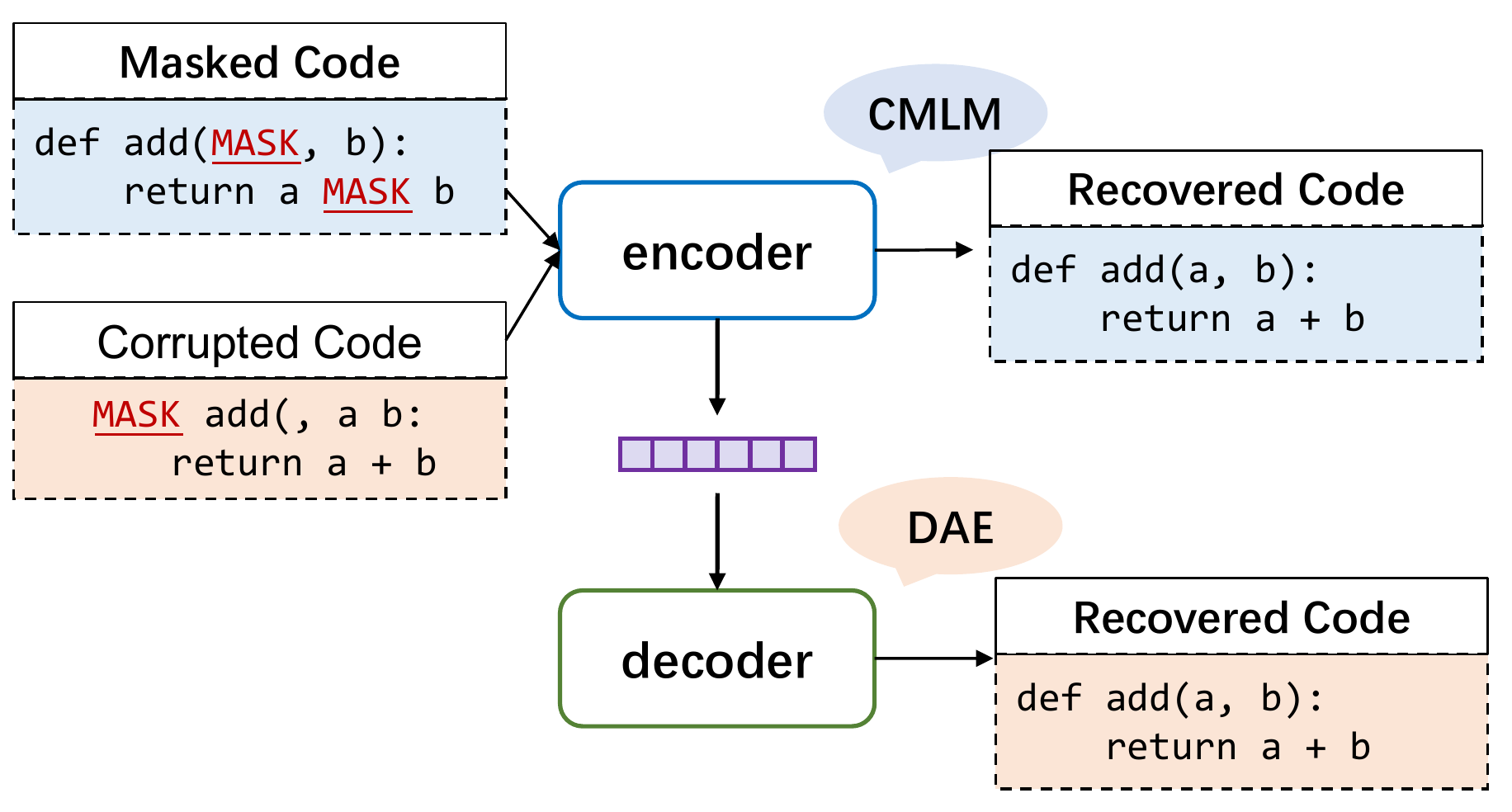}
    \caption{Initialization pre-training.}
    \label{fig:train_1}
    \vspace{-0.3cm}
\end{figure}

\begin{figure}[h]
    \centering
    \includegraphics[width=\linewidth]{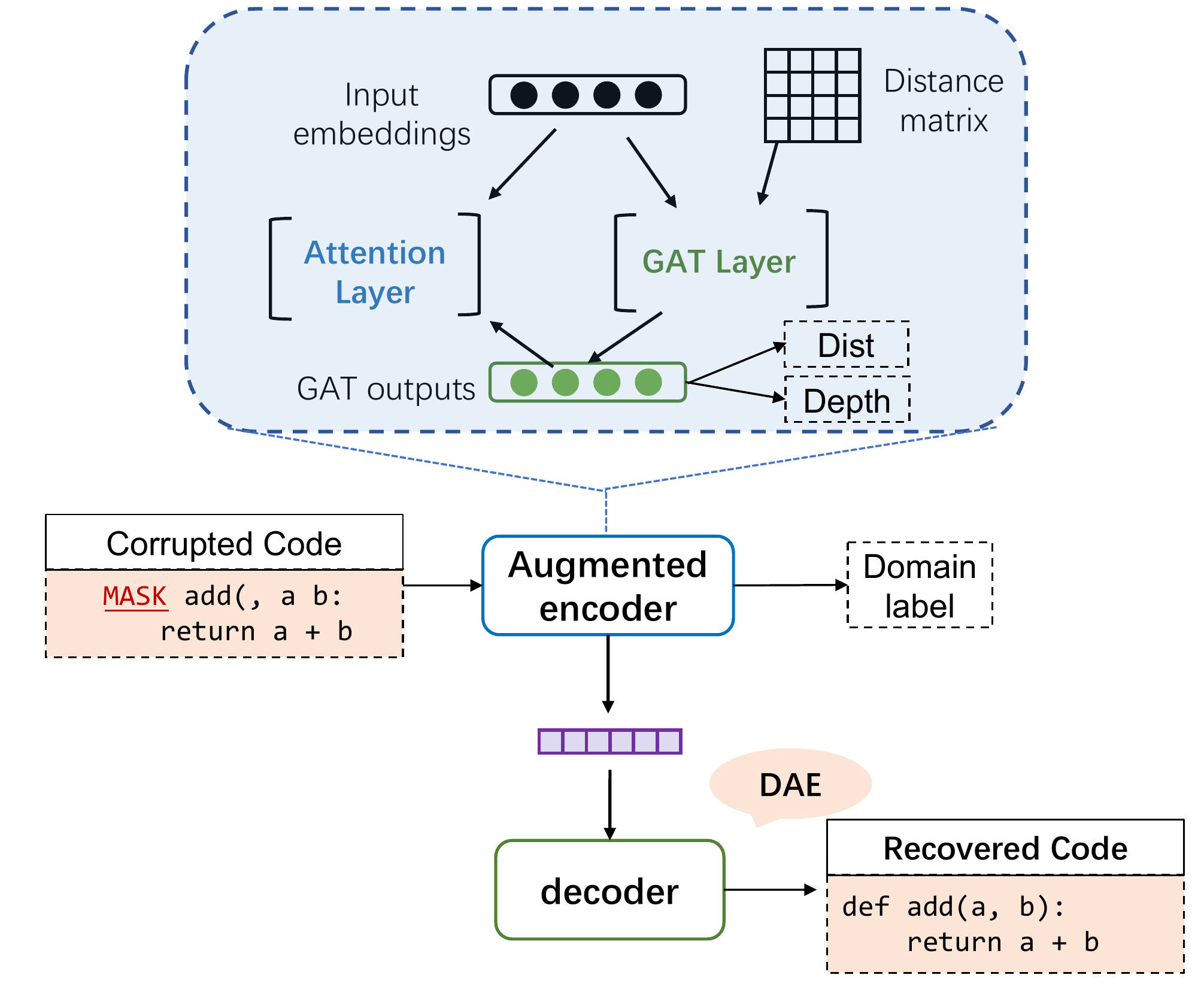}
    \caption{Syntax and domain knowledge augmentation.}
    \label{fig:train_2}
    \vspace{-0.3cm}
\end{figure}

\begin{figure*}[h]
    \centering
    \includegraphics[width=0.8\linewidth]{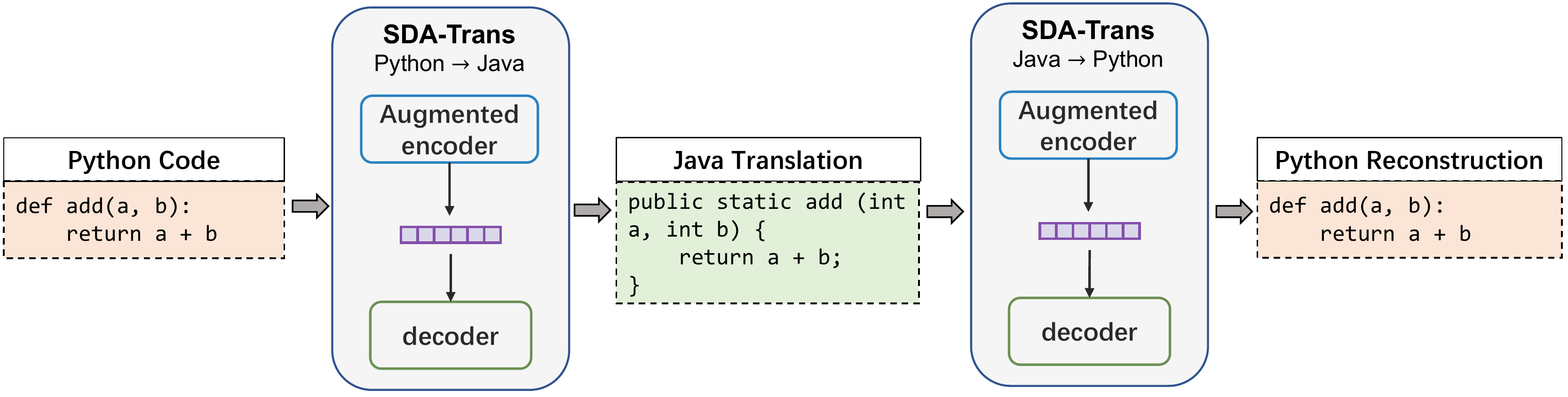}
    \caption{Back-translation tuning.}
    \label{fig:train_3}
    \vspace{-0.3cm}
\end{figure*}

Inspired by the unsupervised machine translation training paradigm, we train SDA-Trans with the following steps: (1) Initializing model parameters with Cross-lingual Masked Language Model and Denoising Auto-Encoding tasks; (2) Augment the model with syntax and domain-related knowledge, and propose several specific pre-training tasks to continuously train the model, making our model syntax and domain aware; (3) Finally use back-translation to teach our model to learn how to translate functions in a weakly-supervised way. The model is shared for all programming languages during these training stages. 
The illustration of the training procedure is shown in Figure \ref{fig:train_1}, \ref{fig:train_2}, and \ref{fig:train_3}.

\subsection{Initialization Pre-training}
To ensure our model learns high-quality cross-lingual token and code sequence representations and further capture the cross-lingual semantics of the code snippets, we pre-train a vanilla Transformer-based seq2seq model with Cross-lingual Masked Language Model (CMLM) and Denoising Auto-Encoding (DAE) objectives on monolingual source code datasets. The parameters of the encoder and decoder are initialized randomly. For the CMLM objective, we randomly mask out some of the tokens in the code snippets and train the encoder to predict these tokens based on their contexts. 
To further teach the decoder to learn how to decode a code sequence based on the encoder's output, in the meantime, we train our model using a Denoising Auto-Encoding (DAE) objective. Specifically, the input code sequence is firstly corrupted with noising functions, then train the model to recover the original code. To corrupt a sequence, we follow \citet{Lample2018UnsupervisedNMT} and \citet{TranCoder2020} to remove, mask, and shuffle input tokens randomly.  
With the DAE objective, the encoder will become more robust for handling the noises, which can benefit the next back-translation stage. Besides, it also helps the decoder learn to generate correct functions \cite{TranCoder2020}. These two objectives are optimized simultaneously.
To offer the model the ability to learn cross-lingual representations and generate code of different programming languages, we follow \citet{TranCoder2020} to alternate between batches of different languages during the training.
CMLM and DAE are illustrated in Figure \ref{fig:train_1}.

\subsection{Continuously Training with Syntax and Domain Augmentation}
After the initialize pre-training, we enhance the model by continuously training with syntax and domain knowledge augmentation using the DAE objective. As described in Section \ref{syntax_augmented_encoder}, we use GAT output representation to bias the self-attention. The parameters of GAT are initialized randomly and are tuned with other parameters jointly. 
It is hard for the DAE training objective to directly guide GAT to learn the syntax structure knowledge. To address this issue, we propose auxiliary objectives to supervise GAT to learn structural representations based on the AST. Specifically, we use GAT’s output representations to predict the distance $D_{ij}$ between token pairs $(s_i,s_j)$ and the tree depth $D_{i}$ of input token $s_i$. The parameters are optimized by minimizing the following objectives:
\begin{equation}
    \begin{split}
        &\mathcal{L}_{dis} = \sum_s \frac{1}{n^2}\sum_{ij} |{D_{ij}}^2-(W_1(f_i-f_j))^T(W_1(f_i-f_j))| \\
        &\mathcal{L}_{depth} = \sum_s \frac{1}{n^2}\sum_{i} |{D_{i}}^2-(W_2(f_i))^T(W_2(f_i))|
    \end{split}
\end{equation}
where $s$ is the input token sequences in the training dataset, $f_i$ is the GAT representation of token $i$, and $W_1$, $W_2$ are the linear projection parameters.

Besides, to encourage the model to extract the language-invariant features, we also design a domain distinguish task. We add a domain discriminator layer to predict the language labels of the source text, and the parameters of our encoder are optimized to maximize the domain discriminator's loss by adversarial training. This objective will encourage the encoder to deceit the domain discriminator and generate language-invariant features. We first perform the mean pooling over the encoder output, then feed the mean value $\tilde{h}$ to the gradient reversal layer (GRL) \cite{ganin2016domain} $\mathcal{Q}_{\lambda}$ before feeding it to the domain discriminator.
\begin{equation}
    d = \text{softmax}(W_d \mathcal{Q}_{\lambda}(\tilde{h}) +b_d
\end{equation}
The objective is to minimize the cross-entropy for all data from different programming languages (domains):
\begin{equation}
    \mathcal{L}_{domain} = - \frac{1}{N_{langs}}\sum_i^N \sum_j\hat{d}^i(j) \text{log}d^i(j) \
\end{equation}
where $\hat{d}^i \in {0,1,...,N_{langs}-1}$ is the ground truth domain label.
Finally, our model is trained by minimizing the loss: $\mathcal{L} = \mathcal{L}_{DAE} + \alpha(\mathcal{L}_{dis} + \mathcal{L}_{depth}) + \beta \mathcal{L}_{domain}$, where $\alpha$ denotes weight for syntax structure prediction loss (distance and depth prediction losses), and $\beta$ is the weight for the domain distinguish loss.

\subsection{Fine-tuning with Back-translation}

Previous DAE training primarily teaches the model to figure out the synthetic disturbance, where the copying operation is the primary task. Besides, the training data is monolingual for each training step.
However, the final goal of SDA-Trans is to translate between function pairs from different languages. To achieve it, we fine-tune our model through back translation \cite{Sennrich2016BackTrans}. As shown in Figure \ref{fig:train_3}, given a function in language \textit{X}, our model first translates it to the other language \textit{Y}, then we train the model to translate \textit{Y} back to the original code snippet \textit{X}. The two reverse translation tasks are trained in parallel until convergence.

\section{Experimental Setup}

\subsection{Training and Evaluation Data}
The unsupervised program translation approach TransCoder {\cite{TranCoder2020}} relies on pre-training with a huge monolingual code corpus, which consists of 168 GB, 352 GB, and 224 GB for C++, Java, and Python data, respectively. To digest the huge data, TransCoder is trained on 32 V100 GPUs for 12 days. It is costly and resource-consuming, and also unrealistic for most research institutes to train a model with such a massive dataset, also unrealistic for us. This is exactly the reason and motivation that we do this research, figuring out how to design a program translation model that can learn sufficient knowledge from limited data. 
Besides, TransCoder only supports translation between language pairs where the model has been pre-trained. For other low-resource languages or languages that are not seen during the pre-training, the performance drops a lot (as demonstrated in RQ3).
Since SDA-Trans can leverage the syntax and domain-related information, we argue that it can learn sufficient knowledge from limited training data, consequently saving computing resources. In our experiments, we use the Python and Java datasets from CodeSearchNet {\cite{husain2019codesearchnet}} and do not use any C++ data to train our model. Our dataset is much smaller than the origin training dataset in TransCoder, and the detailed data statistics are presented in Table {\ref{tab:pretrain_data}}. 

\begin{table}[h]
    \centering
    \setlength{\abovecaptionskip}{0.1cm} 
    \caption{Pretraining data statistics used in TransCoder\cite{TranCoder2020} and SDA-Trans (\# of functions).}
    \begin{tabular}{l|c|c|c}
    \toprule
    ~ & Python & Java & C++ \\
    \midrule
     TransCoder & 217M & 402M & 120M \\
     SDA-Trans & 0.25M & 0.17M &  0 \\
     \bottomrule
    \end{tabular}
    \vspace{-0.3cm}
    \label{tab:pretrain_data}
\end{table}

The evaluation of the semantic correctness of the translated program is important and challenging. Utilizing the unit tests to evaluate the correctness is a feasible solution. However, it is hard and costly to collect parallel program corpus as well as generate reliable unit tests simultaneously.
\citet{TranCoder2020} created a high-quality program translation evaluation datasets, including a group of parallel functions in Python, C++, and Java collected from GeeksforGeeks\footnote{https://practice.geeksforgeeks.org}. GeeksforGeeks is an online platform that includes many coding problems as well as solutions in different programming languages. In this dataset, some of the programs also have corresponding unit tests, which can be used for semantic correctness evaluation. 
To the best of our knowledge, there is no other better evaluation set. Thus we use the GeeksforGeeks to evaluate the semantic correctness of the translated programs.
The detailed data statistics are shown in Table \ref{tab:eval_data}.

\begin{table}[t]
    \centering
    \setlength{\abovecaptionskip}{0.1cm} 
    \caption{Number of functions for validation and test sets.}
    \begin{tabular}{l|c|c|c}
    \toprule
    ~ & Python & Java & C++ \\
    \midrule
    valid & 550 & 550 & 550 \\
     valid w/ unit tests & 237 & 234 & 231 \\
     test & 868 & 868 & 868 \\
     test w/ unit tests & 463 & 481 & 466 \\
     \bottomrule
    \end{tabular}
    \label{tab:eval_data}
\end{table}

For data pre-processing, we perform tokenization and syntax parsing of code snippets using tree-sitter \footnote{https://github.com/tree-sitter/tree-sitter}. Following previous work \cite{TranCoder2020}, we split tokens into sub-word units with the BPE algorithm.

\subsection{Implementation Details}

\subsubsection{Model Configuration}
Following \citet{TranCoder2020}, both the encoder and decoder of our backbone model use a 6-layer transformer with 8 attention heads, 64 head size, and 1024 model size.
For the graph attention network (GAT), the number of GAT heads is set to 1, and the head dimension is 64. The number of GAT layers is equal to the number of encoder layers. The threshold of computing the pair-wise token distance $\sigma$ is set to 10. The initial value of $\alpha$ and $\beta$ (weights of the tree structure and domain prediction losses) are set to 0.05 and then linear decay to 0.01 within 30,000 steps. 

\subsubsection{Training}
The maximum length of code is 512 tokens, and the batch size is set to 16. Following \cite{TranCoder2020}, we alternate between batches of Python and Java during the training procedure. 
We optimize SDA-Trans with the Adam optimizer \cite{kingma2014adam}. The initial learning rate is 5e-5.
SDA-Trans is implemented in PyTorch\footnote{https://pytorch.org/}. It is trained on two V100 GPUs with 32GB memory. The initialization, augmentation, and back-translation process takes approximately 30 hours altogether.

\subsection{Evaluation Metrics}

\textbf{BLEU} score is widely used in the majority of program translation studies \cite{TranCoder2020,TransCoderST2022}, which evaluates the quality of generated functions by measuring the N-gram overlapping between the reference sentence $\hat{y}$ and the translation $y$. 
\begin{equation}
    \text{BLEU-N} = b(y,\hat{y})\cdot \text{exp}(\sum_{i=1}^N\beta_i \log p_i(y,\hat{y}))
\end{equation}
We use BLEU-4 to measure the quality of the translation.
Another strict but simple metric is the \textbf{Exact Match Accuracy (EM Acc)}, which computes the percentage of translations that perfectly match the golden reference. Exact match accuracy is also used as a metric in our experiments. However, these two metrics do not consider the syntactic correctness and semantic equivalence of the programs. When two semantically equivalent programs are implemented in quite different ways, they will have zero EM Acc and a low BLEU score. To address this issue, \citet{TranCoder2020} propose the \textbf{Computational Accuracy} to evaluate whether the translated code can produce the same results as the ground truth when the input data is the same. The translated program will be considered correct only when it generates the same output as the ground truth program for every input.

\section{Empirical Results}
\subsection{RQ1. How effective is SDA-Trans for program translation? }

\underline{\textit{Motivation.}} We investigate whether our proposed approach obtains improvements in program translation when compared with existing advanced baselines.

\underline{\textit{Experimental Setup.}} We compare with the following large scale pre-trained models: (i) TransCoder \cite{TranCoder2020}, (ii) TranCoder*, and (iii) CodeT5 \cite{CodeT52021}. TransCoder \cite{TranCoder2020} is firstly pre-trained on a huge Python, Java, and C++ monolingual program corpus and then evaluated in Python-Java-C++ translation tasks after back-translation tuning. Since we use the same test datasets to evaluate our model, the translation results of their model in the experiments are copied from their paper. CodeT5 uses 6 programming languages in CodeSearchNet \cite{husain2019codesearchnet} and C/CSharp languages from BigQuery, which consists of more than 8 million training instances. Their model has obtained state-of-the-art results on many code-related tasks, including program translation. To compare with CodeT5, we fine-tune it using the valid set in our evaluation parallel datasets in a supervised way. We also re-train TransCoder model using our small-scale Python and Java datasets and named it TranCoder*. The comparison between TranCoder* and our model can demonstrate how many performance improvements SDA-Trans can bring when using the same training data.

\underline{\textit{Results.}} Our model is pre-trained on Java and Python monolingual datasets. We report the translation results between Java and Python in Table \ref{tab:py_java_res}. The translation results on unseen (not pre-trained) language C++ are presented in RQ3. All the translations are generated using greedy decoding (i.e., beam size is 1). TransCoder performs the best in terms of BLEU-4 score since it is trained on a huge data corpus. However, the results of TranCoder* show that, when using the same pre-training datasets as our model, which is much smaller, the performance drops a lot, and is worse than our model in both two translation tasks. 
Since TransCoder* and SDA-Trans have a comparable model size and use the same training data, the poor performance of TransCoder* could be attributed to the simple data processing (feature extraction) and training objectives without syntax constraints or domain knowledge guidance. Thus, it needs much more training data to learn the syntax and semantic information for the program translation task. By extracting and modeling syntax and domain-invariant features explicitly, SDA-Trans can learn sufficient knowledge from the training dataset.
Our model can achieve comparable results with TransCoder and outperforms it on Python to Java translation in terms of EM Acc. When compared with CodeT5, SDA-Trans outperforms it substantially on both two translation tasks. CodeT5 is fine-tuned with hundreds of parallel data pairs in a supervised way. The results suggest that a large-scale pre-trained model like CodeT5 may not behave well in unseen tasks when the dataset used for supervised fine-tuning is not enough.

\begin{table}[t]
    \centering
    \setlength{\abovecaptionskip}{0.1cm} 
    \caption{Results of translation between Python and Java, programs are generated using greedy decoding.}
    \begin{tabular}{l|c c|c c}
    \toprule
    \multirow{2}{*}{Model} & \multicolumn{2}{c|}{Java $\to$ Python} & \multicolumn{2}{c}{Python $\to$ Java} \\
    \cline{2-5}
    ~ & BLEU & EM Acc & BLEU & EM Acc \\
    \midrule
    TransCoder & 68.1 & 3.7 & 64.6 & 0.8  \\
    CodeT5 &  42.9 & 1.3 & 53.4 & 1.2   \\
    TransCoder* & 57.0 & 3.3 & 49.8  & 0.3   \\
    SDA-Trans & 66.8 & 3.6 & 64.0 & 1.4  \\
    \bottomrule
    \end{tabular}
    \vspace{-0.3cm}
    \label{tab:py_java_res}
\end{table}

\begin{table}[t]
    \centering
    \setlength{\abovecaptionskip}{0.1cm} 
    \caption{Performance of removing syntax augmented or domain aware pre-training from SDA-Trans.}
    \begin{tabular}{l|c c|c c}
    \toprule
    \multirow{2}{*}{Model} & \multicolumn{2}{c|}{Java $\to$ Python} & \multicolumn{2}{c}{Python $\to$ Java} \\
    \cline{2-5}
    ~ & BLEU & EM Acc & BLEU & EM Acc \\
    \midrule
    SDA-Trans & 66.8 & 3.6 & 64.0 & 1.4  \\
    w/o domain & 61.2 & 3.3 & 60.4 & 1.2  \\
    w/o Syn. & 58.5 & 2.5 & 54.4 & 0.7  \\
    \bottomrule
    \end{tabular}
    \vspace{-0.3cm}
    \label{tab:ablation}
\end{table}

\begin{table*}[t]
    \centering
    \setlength{\abovecaptionskip}{0.1cm} 
    \caption{Results of C++ related translation, programs are generated using greedy decoding.}
    \begin{tabular}{l|c c|c c|c c|c c}
    \toprule
    \multirow{2}{*}{Model} & \multicolumn{2}{c|}{Java $\to$ C++} & \multicolumn{2}{c|}{C++ $\to$ Java} & \multicolumn{2}{c|}{Python $\to$ C++} & \multicolumn{2}{c}{C++ $\to$ Python}\\
    \cline{2-9}
    ~ & BLEU & EM Acc & BLEU & EM Acc & BLEU & EM Acc & BLEU & EM Acc \\
    \midrule
    TransCoder & 97.0 & 24.7 & 85.4 & 3.1 & 65.4 & 4.9 & 70.1 & 6.7 \\
    CodeT5 & 72.4 & 11.1 & 60.6 & 5.2 &  32.9 & 0.1 & 28.1 & 0.3\\
    TransCoder* & 74.7 & 14.4 & 71.8 & 11.0 & 50.2 & 1.2 & 49.5 & 2.9\\
    SDA-Trans & 83.4 & 21.8 & 77.7 & 15.6 & 61.9 & 4.2 & 60.8 & 4.1 \\
    \bottomrule
    \end{tabular}
    \vspace{-0.3cm}
    \label{tab:cpp_res}
\end{table*}

\subsection{RQ2. How do different components in SDA-Trans contribute to program translation? }

\underline{\textit{Motivation.}} In this research question, we study whether our proposed components (syntax augmentation and domain-aware pre-training) contribute to the performance improvements of SDA-Trans, and how they impact the results. 

\underline{\textit{Experimental Setup.}} We conduct an ablation study by removing these components from SDA-Trans, respectively, for both pre-training and fine-tuning procedures. When removing syntax augmentation, the encoder turns back to the vanilla Transformer encoder, where GAT and two structure prediction tasks are no longer used. When removing domain-aware pre-training, we remove the domain discriminator layer and do not use the domain-distinguishing task during pre-training.

\underline{\textit{Results.}} The results are presented in Table \ref{tab:ablation}. The first row presents the results of our full model, the second row and the last row present the results of removing the domain distinguish task and syntax augmentation from the full model, respectively. As seen from the results, the model's performance drops when removing each task, and the effect of the syntax augmentation is bigger. Augmenting the model explicitly with the syntax information of the code snippets can encourage the model to distill high-level semantic features instead of focusing on the morphology information, thus can boost the performance of cross-lingual transfer as well as program translation. The domain-distinguishing task is designed and trained (with adversarial training) to encourage SDA-Trans to extract language-invariant semantic features, thus can help to transfer knowledge among different languages.

\subsection{RQ3. How well does our approach perform in unseen language translation scenarios? }

\underline{\textit{Motivation.}}
In this RQ, we aim to examine whether augmenting Transformer with syntax and domain knowledge enhances the cross-lingual transfer ability and how our model performs in unseen language translation scenarios. 

\underline{\textit{Experimental Setup.}} We conduct experiments on the translation between C++ and Python, C++ and Java. The comparison baselines and the model setup are the same as RQ1. Note that C++ language is not seen during the pre-training procedure of SDA-Trans, CodeT5, and TransCoder*. For CodeT5, we fine-tune it using the validation parallel data pairs of our evaluation dataset in a supervised manner. For SDA-Trans and TransCoder*, we apply back-translation in a weakly-supervised way. For TransCoder, we directly copy the results from their paper.

\underline{\textit{Results.}} The results are shown in Table \ref{tab:cpp_res}. As seen from the results, since the training corpus of the TransCoder includes C++ language, it achieves the best performance for almost all translation tasks, except for the EM Acc of C++ $\to$ Java translation. For the other three models, C++ language is never seen during their training procedure. Among these models, SDA-Trans performs best on all the translation tasks. Note that our model even outperforms TransCoder by a large margin in the EM Acc of C++ $\to$ Java translation. It is also worth noting that translation between Java-C++ is easier than Python-C++, as both Java and C++ are object-oriented languages, and the syntax and code morphology are similar. On the contrary, the syntax and code implementations between Python and C++ vary widely. This can also be reflected in the higher BLEU and EM Acc scores for translation tasks between Java and C++. Both CodeT5 and TranCoder* perform badly in the translation tasks between Python and C++. Compared to these two models, the improvements of SDA-Trans are more significant in Python-C++ translation than that in Java-C++. The above results demonstrate that augmenting the model with syntax and domain knowledge enhances the cross-lingual transfer ability of SDA-Trans, leading to great performance in unseen language translation scenarios.
Among all the models, CodeT5 behaves the worst, which further demonstrates that the supervised fine-tuning using small-scale unseen languages does not work well.

\begin{table*}[t]
    \centering
    \setlength{\abovecaptionskip}{0.1cm} 
    \caption{Computational accuracy results of translation between Python, Java, and C++, programs are generated with the beam size 5.}
    \begin{tabular}{l|c|c|c|c|c|c}
    \toprule
     ~ & Java $\to$ Python & Python $\to$ Java & Java $\to$ C++ & C++ $\to$ Java & Python $\to$ C++ & C++ $\to$ Python \\
    \midrule
    Rule-based & 38.3 & - & - & 61.0 & - & - \\
    TransCoder & 60.0 & 44.3 & 86.9 & 70.0 & 44.4 & 58.3  \\
    CodeT5 & 7.6 & 5.8 & 25.2 & 19.9 & 6.8 & 6.1 \\
    TransCoder* & 44.8 & 30.0 & 61.5 & 49.1 & 20.4 & 26.6 \\
    SDA-Trans & 58.1 & 42.6 & 84.7 & 66.3 & 41.2 & 55.4  \\
    \bottomrule
    \end{tabular}
    \vspace{-0.3cm}
    \label{tab:ca_res}
\end{table*}

\subsection{RQ4. How useful are the programs translated by SDA-Trans in practice?}

\underline{\textit{Motivation.}}
In this RQ, we aim to analyze the semantic correctness and usefulness of the programs translated by SDA-Trans. 

\underline{\textit{Experimental Setup.}} 
Since BLEU-4 and EM Acc metrics may not directly measure the semantic correctness of the results, we use computational accuracy \cite{TranCoder2020} to evaluate whether the programs produced by SDA-Trans can generate the same outputs as the ground truth given the same input data. Except for previous baselines, we also compare the results with previously introduced baseline approaches and two rule-based approaches:
j2py\footnote{https://github.com/natural/java2python} for Java $\to$ Python translation, and a commercial tool\footnote{https://www.tangiblesoftwaresolutions.com/}, which can perform translation from C++ to Java. In this experiment, all the translations are generated using beam size 5. 

\underline{\textit{Results.}}
The results are shown in \ref{tab:ca_res}. 
As seen from the results, SDA-Trans achieves comparable computational accuracy scores with TransCoder, and outperforms other approaches by a large margin. The computational accuracy score gap between SDA-Trans and TransCoder is much smaller than the BLEU-4 and EM Acc scores, demonstrating that most of the programs generated by SDA-Trans are semantically equivalent to the reference programs, and can successfully pass the unit tests, even though it looks less similar than the reference compared with the generated programs of TransCoder.

\section{Discussion}

\subsection{Visualization of Features}
\begin{figure}[t]
    \setlength{\abovecaptionskip}{0.1cm} 
    \centering
    \includegraphics[width=\linewidth]{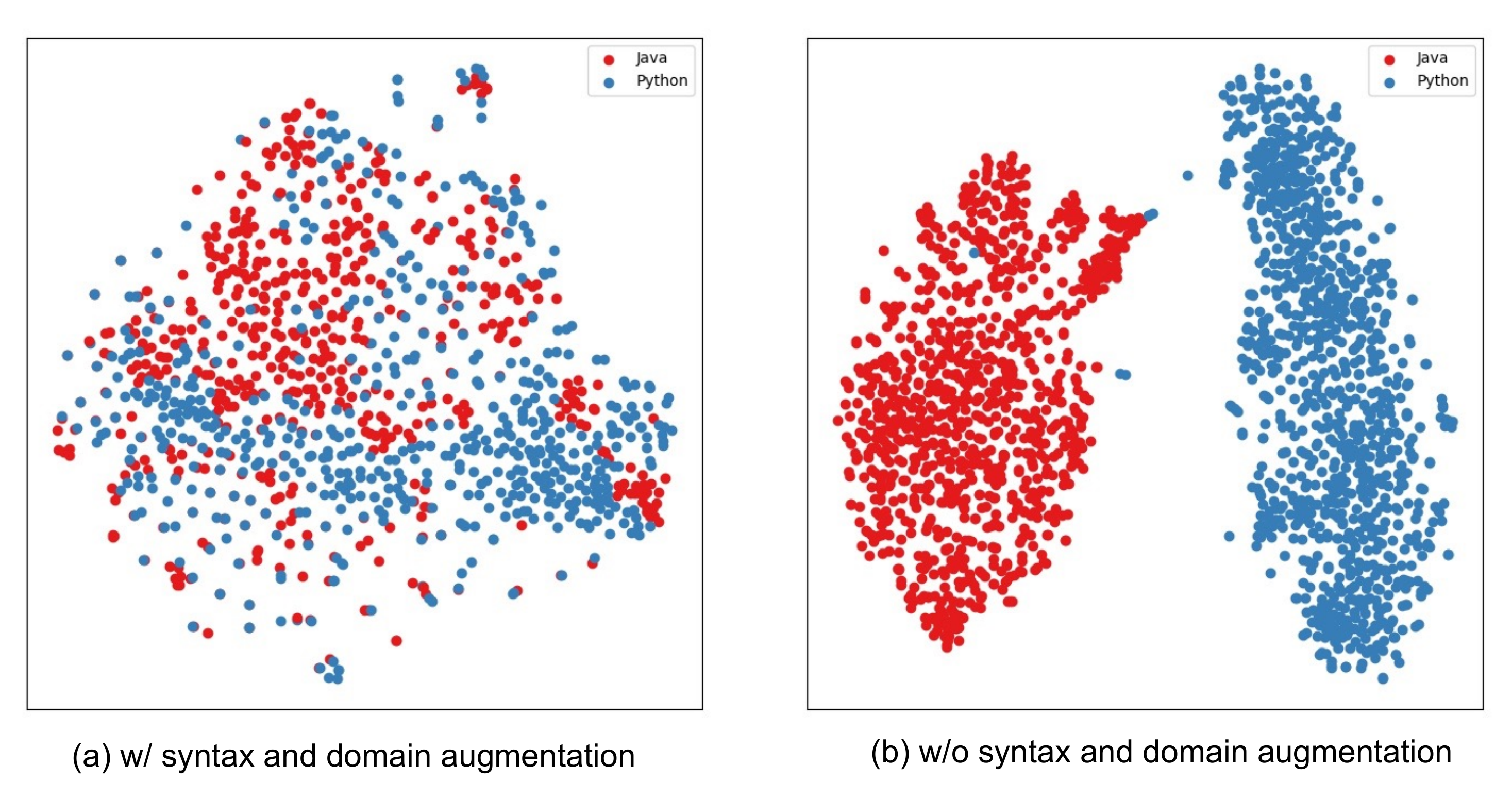}
    \caption{Visualization of the encoder output features.}
    \label{fig:tsne}
    \vspace{-0.3cm}
\end{figure}

We further perform a visualization of the encoder feature representations of the variants of SDA-Trans for the Python-Java translation task. As shown in Figure \ref{fig:tsne}, the graphs are obtained by applying t-SNE \cite{van2008visualizing} on the representations of source data points $\hat{h}$ for both Python $\to$ Java and Java $\to$ Python translation tasks. Every sample is first mapped into a 1024-dimensional vector through the encoder and the mean-pooling operation, and then we projected the vector into a two-dimensional plane by the t-SNE.

Figure \ref{fig:tsne} (a) shows the representation produced by our syntax and domain knowledge augmentation encoder, and Figure \ref{fig:tsne} (b) presents the representation produced by the vanilla Transformer encoder. We can observe that the data points of different languages are well separated in Figure \ref{fig:tsne} (b), which shows that the vanilla Transformer encoder is sensitive to the language type. The data points of Python and Java source functions in Figure \ref{fig:tsne} (a) are mixed together. That is, data samples from different domains/languages are very close to each other. This further suggests that through domain-distinguishing adversarial training and syntax structure modeling, SDA-Trans can distill more complicated domain-invariant features, thus can enhance the cross-lingual transfer ability.

\subsection{Threats to Validity}
\noindent\textbf{Threats to internal validity} include the influence of the model hyper-parameter settings for both our model and the reproduced baseline model TransCoder*. 
To ensure a fair comparison between TransCoder and SDA-Trans, we use the same setting with the strong baseline TransCoder {\cite{TranCoder2020}} in our backbone model. For our proposed components, i.e., GAT layers and heads, we chose the hyper-parameters through a small-scale grid search and manual selection. 
For TransCoder*, as the authors of {\cite{TranCoder2020}} offered the artifact and running scripts to re-train their model with new training data \footnote{https://github.com/facebookresearch/TransCoder\#train-a-new-model}. We follow their scripts to train TranCoder* with our data. During the training, we tuned the mini-batch size and initial learning rate according to our GPU memory and the model's initial performance. Then we use the same configuration in SDA-Trans. We have tried our best to make sure the training process is stable and finish training until the model is converged. However, it cannot be denied that when the datasets changes, the original model's performance could be affected. We have tried our best to make TransCoder* approximate its optimal performance, and make the comparison between TransCoder* and SDA-Trans fair. Therefore, there is a minor threat to the hyper-parameter tuning.

\noindent\textbf{Threats to external validity} include the quality of the datasets. We use the Python and Java programs of the CodeSearchNet dataset \cite{husain2019codesearchnet} to pre-train our model. CodeSearchNet contains function-document pairs collected from GitHub repositories, covering 6 kinds of languages. It is widely used in code pre-train models \cite{codebert2020,CodeT52021}, and these model has achieved great performance in various downstream tasks. Thus, most of the programs can be viewed as high-quality. For evaluation, we use the dataset built in TransCoder \cite{TranCoder2020}. The evaluation set is composed of hundreds of parallel functions in 3 languages, and some of the programs have unit tests that can be used to evaluate the correctness of generated translations. Nerveless, the evaluation scale is not big and comprehensive enough, and further evaluation on big data and other programming languages is needed.
Besides, a further threat is related to the baseline choice. TransCoder-ST {\cite{TransCoderST2022}} is state-of-the-art for unsupervised code translation, which outperforms TransCoder {\cite{TranCoder2020}} substantially. It is an incremental work on TransCoder and relies on large-scale pre-trained models as its backbone. 
Specifically, TransCoder-ST improves TransCoder by filtering out invalid translations through an automated unit-testing system during back-translation, which reduces the noise and further boosts translation performance. The essential question of high training costs and poor cross-lingual transfer is not solved. We believe their filtering technology is orthogonal to SDA-Trans, which can also be used to improve the back-translation performance in our model. Thus, we don't consider it as a baseline in this paper.

\noindent\textbf{Threats to construct validity} relate to the property of the evaluation measure. We adopted BLEU-4 score and exact match accuracy to measure the quality of the translated programs by comparing the predicted programs with the ground truth, and these two metrics are generally used in the previous program translation work. Besides, we further introduce computational accuracy to evaluate the semantic correctness of the generated translations.

\section{Related Work}

\subsection{Deep Learning for Programming Languages}
In recent years, deep learning technologies have been widely used software engineering tasks, such as code completion \cite{liu2020cuglm,svyatkovskiy2019pythia}, code summarization \cite{Hu2018deepcom,Ahmad2020TransformerCS}, code search \cite{Gu2018DCS,Cambronero2019DLcodesearch}, bug detection and repair \cite{Wang2018repair,Yasunaga2020repair}. Unsupervised pre-training methods for code also attracted wide attention and have achieved impressive performance on many downstream tasks. CodeBERT \cite{codebert2020} and GraphCodeBERT \cite{graphcodebert2021} are encoder-only models that are based on BERT \cite{BERT2019}.  PLBART \cite{PLBART2021} is pre-trained as a generative denoising autoencoder based on BART \cite{BART2020}. 
DOBF is a deobfuscation-based pre-train model {\cite{DOBF2021}}. It designs a deobfuscation pre-training objective, which adds noise based on the code structure. Specifically, it obfuscates code snippets by replacing names of classes, functions, and variables with special symbols, and trains their model to recover their initial names. DOBF outperforms GraphCodeBERT on multiple code-related tasks.
CodeT5 \cite{CodeT52021} is an encoder-decoder architecture based on T5 \cite{T52019}, which proposes two code semantic aware pre-training objectives. It achieves advanced performance at both code understanding and generation tasks. 

\subsection{Structure Learning for Programming Language Modeling}
Language syntax structural information has been proved effective for representation learning and cross-lingual transfer in both natural language and programming language \cite{xie2020contextual,ahmad2021gate,Ahmad2020SynBERT,codetransformer2021,graphcodebert2021}. 
\citet{Ahmad2020SynBERT} proposed to augment BERT \cite{BERT2019} with a universal dependency tree structure while fine-tuning on downstream tasks. \citet{ahmad2021gate} explicitly fused structural information by incorporating different syntactic distances into the self-attention mechanism. It achieved great performance on cross-lingual relation and event extraction tasks.
In recent years, incorporating code structure into neural networks has become a hot topic, and various models have also been proposed.
Code Transformer \cite{codetransformer2021} considered several semantic distances when operating self-attention, covering both structural and contextual information. It achieved strong performance gains in low-resource languages. GraphCodeBERT \cite{graphcodebert2021} is a pre-trained model, which injects the data-flow knowledge of the code snippet into the code representation by two structure-aware pre-training tasks. 

\subsection{Translation of Programming Languages}
\noindent\textbf{Supervised NMT-based Approaches.}~ With the development of NMT techniques, researchers began to explore the possibility of adapting neural machine translation to program translation task. \citet{chen2018tree} proposed a tree-to-tree neural architecture. They parsed programs into ASTs and converted them into binary trees, then fed the trees into a Tree-LSTM based encoder-decoder neural model. \citet{gu2017deepam} proposed DeepAM based on an RNN sequence-to-sequence model, which automatically diggings API mappings between programming language pairs. In recent years, pre-trained models have achieved remarkable performance on both NLP \cite{BERT2019,gpt-2} and SE \cite{codebert2020,CodeT52021,liu2020cuglm} fields. Some approaches have been applied for program translation tasks. CodeT5 \cite{CodeT52021} is a unified encoder-decoder Transformer model and has achieved superior performance on many code understanding and generation downstream tasks, including program translation.

\noindent\textbf{Unsupervised NMT-based Approaches.}~ 
NMT-based models heavily rely on parallel data. However, parallel resources are costly to obtain. Several studies try to adapt unsupervised machine translation approaches for program translation. TransCoder \cite{TranCoder2020} first proposed to apply unsupervised machine translation approaches for program translation task, which is trained with a huge monolingual code corpus. Their model can be applied for translating between C++, Java, and Python. Later, TransCoder-ST \cite{TransCoderST2022} improves TransCoder by filtering out invalid translations through an automated unit-testing system during back-translation, which reduces the noise and further boosts the translation performance.

\section{Conclusion and Future Work}
In this paper, we propose SDA-Trans, a syntax and domain aware model for unsupervised program translation. Specifically, we augment the transformer by incorporating syntax information into the self-attention mechanism, and design a new domain-distinguish pre-training task to distill the domain-invariant features by adversarial training. Our model is pre-trained using a combination of cross-lingual masked language modeling and denoising objectives using a smaller-scale monolingual Python and Java corpora, and finally tuned with back-translation. Experimental results on translation tasks between Python, Java, and C++ demonstrate that SDA-Trans matches or outperforms the large-scale pre-trained methods, especially on unseen C++ language translation.

\section*{Acknowledgment}
This work is funded by the National Science Foundation of China Grant No. 62177003. It is also supported by the State Key Laboratory of Software Development Environment No. SKLSDE-2022ZX-13.

\bibliography{ref}

\end{document}